\documentclass[
aps,%
12pt,%
final,%
notitlepage,%
oneside,%
onecolumn,%
nobibnotes,%
nofootinbib,%
superscriptaddress,%
noshowpacs,%
centertags]%
{revtex4}

\begin{document}
\selectlanguage{english}

\title{On The Existence of Planets Around the Pulsar PSR B0329+54}

\author{\firstname{E.~D.}~\surname{Starovoit}}Starovoit E.D. 
\email{starovoyt.prao@gmail.com}
\affiliation{Pushchino Radio Astronomy Observatory, Astro Space Center, P.N. Lebedev Physical Institute, Russian Academy of Sciences, Pushchino, Moscow region, Russia}
\affiliation{Pushchino State Natural Sciences Institute, Pushchino, Moscow region, Russia}
\author{\firstname{A.~E.}~\surname{Rodin}}Rodin A.E.
\email{rodin@prao.ru}
\affiliation{Pushchino Radio Astronomy Observatory, Astro Space Center, P.N. Lebedev Physical Institute, Russian Academy of Sciences, Pushchino, Moscow region, Russia}

\begin{abstract}
Results of timing measurements of the pulsar PSR B0329+54 obtained in 1968--2012 using
the Big Scanning Antenna of the Pushchino Radio Astronomy Observatory (at 102 and 111 MHz), the
DSS 13 and DSS 14 telescopes of the Jet Propulsion Laboratory (2388 MHz), and the 64 m telescope
of the Kalyazin Radio Astronomy Observatory (610 MHz) are presented. The astrometric and rotational
parameters of the pulsar are derived at a new epoch. Periodic variations in the barycentric timing residuals
have been found, which can be explained by the presence of a planet orbiting the pulsar, with an orbital period $P_{1}$ = 27.8 yr, mass \textit{$m_{c}$}sin\textit{i} = 2$M_{\oplus}$, and orbital semi-major axis $a$ = 10.26 AU. The results of this study do not confirm existence of a proposed second planet with orbital period $P_{2}$ = 3 yr.

\end{abstract}

\maketitle

\section{INTRODUCTION}

The pulsar B0329+54 was discovered in 1968 by Cole and Pilkington \cite{cole1968}. In 1979, Demia\'{n}ski and Pr\'{o}szy\'{n}ski \cite{demianski1979} suggested that the pulsar may have a planet with a period of three years. In 1985, Cordes and Downs \cite{cordes1985} did not confirm this periodicity, while
Bailes et al. \cite{bailes1993} again found this periodicity in data
obtained with the 76 m radio telescope of the Jodrell
Bank Observatory.

In 1995, based on newly analyzed data from the
Jet Propulsion Laboratory (JPL) \cite{cordes1985} and her own data
obtained with the Big Scanning Antenna (BSA)
of the Pushchino Radio Astronomy Observatory
(PRAO), Shabanova \cite{shab1} concluded that the timing
data does contain a three-year periodicity, and that
variations with a period of about 17 yrs are also
present.
 
In 1999, Konacki et al. \cite{konaki} again turned to the
question of planets. Based on observations with the
100-m Effelsberg radio telescope (Germany) and the
32-m Toru\'{n} radio telescope (Poland), they concluded
that the new timing data are not described well by the
earlier model with two planets, implying that there are
no planets around the pulsar.
 
Hobbs et al. \cite{hobbs1} present a timing-residual plot
that shows the presence of quasi-harmonic variations
covering two periods of a possible planetary companion.
However, it was concluded that the variations
are most likely due to the timing noise of the pulsar itself. Note that Shabanova et al. \cite{shab2} agree with the
conclusions of Hobbs et al. \cite{hobbs1} regarding the absence
of planets near this pulsar.

As can be seen from the above references, the
question of the existence of planets orbiting
PSR B0329+54 remains open. We therefore decided
to perform a detailed study based on all currently
available data (the pulsar has been observed since
1968).

\section{OBSERVATIONS}

We used data obtained during 1968--2012, at three
radio observatories and at three different frequencies:
\begin{enumerate}
\item 1968--1985 -- the DSS-13 (26-m diameter)
and DSS 14 (64-m diameter) radio telescope of the
JPL, operating at 2388 MHz. The observing bands
were 12 and 32MHz \cite{downs}, and the time spanned by the
observations was MJD = 40 105--45 385.
\item 1978--2012 -- the PRAO BSA, operating at
102 and 111 MHz. The observing band was 2.5 MHz \cite{shab2}, and the time spanned by the observations
was MJD = 43 702--55 973.
\item 1997--2001 -- the 64-m radio telescope of the
Kalyazin Radio Astronomy Observatory (KRAO) operating
at 610 MHz. The observing band was 2 $\times$ 3.2 MHz \cite{Ilyasov}, and the time spanned by the observations
was MJD = 50 645--52 264.
\end{enumerate}

In all, the observations we used for timing of
PSR B0329+54 cover 44 yrs, and are still ongoing.

From the start of observations with the LPA, its
operating range had been 101--104 MHz, but it was readjusted in 1998 to observe at 109--113 MHz.
Therefore, PSR B0329+54 was observed with this
instrument at 102 MHz during 1978--1998, and at
111 MHz from 1998 to the present. A 64-channel
radiometer was used for these observations. The
channel bandwidth was 20 kHz; a 512-channel
spectrum analyzer was used starting in 2005 \cite{logvinenko}.

Observations of PSR B0329+54 at JPL were
made at 2388 MHz, starting in 1968 with the 26-meter DSS-13 antenna with a bandwidth of 12MHz,
and later with the 64-meter DSS-14 antenna with a
bandwidth of 32 MHz \cite{downs}.  

Pulsar observations at the KRAO were made using
the fully steerable 64-m radio telescope at an
operating frequency of 610 MHz \cite{Ilyasov}. 

All the recording systems applied pulse accumulation
synchronous with the apparent period. In each
case, the number of combined pulses was chosen
individually, depending on the sampling interval and
the width of the frequency channel, in order to guarantee
a high signal-to-noise ratio. The accumulation
interval for the BSA was determined by the time for
the passage of the pulsar through the antenna beam. 

In all, 543 pulse arrival times (PATs) were obtained
at JPL, 229 at the KRAO, and 3037 at the PRAO.

\section{DATA REDUCTION}

All the timing data were processed using the
Tempo software package.\footnote{http://tempo.sourceforge.net/} 
The model for the phase
analysis of the PATs included the rotational (the
frequency and its derivatives) and astrometric (the
coordinates and proper motion) parameters of the
pulsar. The delay model is described in detail by
Doroshenko and Kopeikin \cite{doroshenko}, as well as Hobbs
et al. \cite{hobbs2,edwards}. The analysis of the behavior of the
residuals in the presence of errors in the parameters
determined is also described in these same papers.
When processing the entire dataset for the three
observatories, it was found \cite{rs2017} that the coordinates
of the pulsar cannot be determined simultaneously at
different frequencies. Therefore, the position of the
pulsar on the sky was first determined separately for
the three datasets, after which the timing residuals
were reduced to a single plot with common spin
parameters.

To minimize the influence of the errors in the pulsar
proper motion, its coordinates were reduced to epoch
MJD = 48 000, which is close to the middle of the
entire time interval spanned by the observations. Table
1 presents the astrometric and spin parameters of
the pulsar.

The analysis performed using the Tempo software
yielded timing residuals for PSR B0329+54 after fitting
with a cubic polynomial covering the entire time
span of the observations. The result is plotted in
Figure 1. 

As can be seen, the timing residuals indicate the
presence of a quasi-harmonic modulation, which
suggests the existence of a planet orbiting the pulsar.
The plot also shows that the data form a double \textquotedblleft noise
corridor\textquotedblright, due to the fact that PSR B0329+54 is a
mode-switching pulsar. When the mode switches,
the shape of the pulse changes and it is simultaneously
delayed by approximately 1 ms \cite{shabanova1994}. We did not
take this into account in our present study, and have
not analyzed it separately.

We fitted a harmonic curve describing the effect
of a possible planet to the residuals, shown in Fig. 2
by the solid curve. The period of the harmonic variations
is 28 yrs. After excluding the 28-yr periodicity
and variations in the rotation of the pulsar itself, the
other, so-called three-year, periodicity first noted by
Demia\'{n}ski and Pr\'{o}szy\'{n}ski \cite{demianski1979} and then confirmed by
Shabanova \cite{shab1} becomes clearly visible (see Fig. 3).
Figure 3 clearly shows that, at the beginning of the
observations, in 1968--1978, the variations had the
period of 3 yrs and an amplitude of 1 ms, while the
character of these variations has changed significantly
since the 1990s: the amplitude increased to
1.6 ms and the period increased to 4.4 yr. Such a
change in the variations of the residuals is extremely difficult to explain using a model with a planet; this
would require the transition of the planet into another,
longer-period, orbit under the action of some external
force.

\section{RESULTS}

The question of the existence of a planet is complicated
by the fact that a number of pulsars display
peculiarities in their rotation, which, under certain conditions, can produce quasi-periodic variations of
the PATs, resembling the influence of an external body
orbiting the pulsar. The opposite situation is also
possible: strictly harmonic deviations of PATs due
to a planet can be completely distorted by random
variations of the pulsar spin phase, leading to the
incorrect conclusion that there is no companion in
orbit around the pulsar. Therefore, strictly speaking,
the time span of the observations, which covers only
two to three orbital periods, is not sufficient to enable
unambiguous conclusions about the presence or absence
of a planet around the pulsar. Confirmation of
the existence of a planet would requires longer series
of observations.

Observational studies of binary systems give the
semi-major axis of the pulsar orbit projected onto the
line of sight and the orbital period, listed in Table 2.
These quantities enable calculation of the mass function,
\begin{equation}
f(m_p)=\frac{(m_c\sin i)^3}{m_p+m_c}=\frac{4\pi^2}{G}\frac{(a_p\sin i)^3}{P_b^2}\simeq 1.04\cdot 10^{-16}M_\odot.
\end{equation} 

If we suppose the pulsar mass to be $m_p=1.44M_\odot$, the mass of the planet is two Earth masses: $m_c\sin i \simeq 2M_\oplus$. The semi-major axis of the relative orbit derived from Kepler's third law is
\begin{equation}
a=\sqrt[3]{m_p P_b^2}\simeq 10\;{\rm AU}.
\end{equation}

The calculated parameters of the planet are presented
in Table 2.

We cannot currently confirm the existence of a
second planet around PSR B0329+54, with an orbital
period much shorter than the period of the first
planet. The presence of short-period variations with
the period 3--4.4 yrs is described well by a sawtoothlike
curve, whose nature indicates the action of a
mechanism inside or near the pulsar that leads to discontinuous
variations of the pulsar's rate of rotation,
with the character of these processes changing after
about 20 yrs of timing measurements. This, in turn,
could be due, for instance, to different rates of braking
of the pulsar due to changes in the energy-loss rate
or the periodic reorganization of the internal structure
of the neutron star. Firmer conclusions about the
origin of the variations in the pulsar's rotation rate
will become possible only after detailed studies of the
evolution of the pulse shape, which is the subject of a
dedicated future study.

However, the possible existence of a second planet
should not be completely lost from sight. Continued
observations of PSR B0329+54 are necessary,
and additional data may make it possible to finally
determine whether this neutron star has a second
planet, or what processes give rise to the appearance
of quasi-periodicity in the currently observed post-fit
residuals. In addition, the amount of observational
data currently available is not sufficient to fully cover
at least two orbital periods of the first, long-period
planet.

\section{DISCUSSION}

Several groups of researchers are currently observing
hundreds of pulsars, but \textquotedblleft pulsar\textquotedblright{} planets
remain a fairly rare phenomenon against the background
of the thousands of planets discovered around
stars in other stages of their evolution. This raises the
question of the origin of pulsar planets, in particular,
of the planet orbiting PSR B0329+54.

At the moment, there are two models for the origin
of pulsar planets.
\begin{enumerate}
\item The planet orbiting the neutron star formed
after the supernova explosion, from stellar matter that
remained in orbit.
\item The planet formed before the pulsar, survived
the supernova explosion, and remained in orbit
around the pulsar.
\end{enumerate}

Each of these theories has its own properties and
conditions.

If the planet forms after the supernova, there
must remain after the explosion, formation of the
neutron star, and ejection of the stellar envelope
enough matter for the formation of this new body,
or even of a planetary system, as in the case of
PSR B1257+12 \cite{wolszczan}. Further, this condition must
be met in the presence of the huge expansion velocity
of the ejected shell ($10^3-10^4$ km/s) in the case of a
symmetric explosion, or in the case of the motion of
the neutron star through the expanding supernova
shell, if the explosion was asymmetric.

In the second model, the planet ends up during
the supernova explosion in the path of an ejected
shell with a mass of several solar masses propagating
with a velocity of $10^{3}-10^{4}$
km/s. The destruction
of the planetary system is mainly brought about
by the pressure of the ejected stellar matter, which
can exceed the gravitational attraction between the
planets and the central star and push the planet or
planets out of the stellar system; all this is accompanied
by a significant decrease of the mass of the
central body. In addition, even if the planet remains in orbit around the star, a solid planet will lose its
atmosphere, which will be carried away by the ejected
stellar shell, while a gaseous planet will lose some
of its mass. Even if the planet survives the supernova
explosion, the semi-major axis and eccentricity
of its orbit should increase. Thus, the almost perfectly
circular orbits of the planets around the pulsar
PSR B1257+12 \cite{wolszczan} are not consistent with this
theory. However, if we take the orbital eccentricity
of the planet around PSR B0329+54 with the period $P_{1}$ = 27.8 yrs to be \textit{$e_{1}$} = 0.236, we can assume that
the orbit of this planetary companion to the neutron
star became elliptical in the process of the supernova
explosion.

Answering the question of whether a planet can
survive the evolutionary transformation from a regular
star to a neutron star, while continuing to orbit this
object, requires the construction of a model for the
evolution of the planetary orbit during the supernova
explosion. This model must take into account all
perturbing forces hindering the preservation of the
planetary system, and consider models for both symmetric
and asymmetric explosions. We plan to explore
this in future studies.

\section{CONCLUSION}

The main results of this study are the following.
\begin{enumerate}
\item Timing of the pulsar PSR B0329+54 was carried
out in 1968--2012 using observations obtained
with the PRAO LPA (102, 111 MHz), the KRAO 64-m radio telescope (610 MHz), and the JPL DSS-13 and DSS-14 telescopes (2388 MHz). We have
derived the astrometric and spin parameters of the
pulsar at epochMJD = 48 000. 
\item Quasi-periodic variations that are described
well by a model with a planet having an orbital period $P_{1}$ = 27.8 yrs were detected in the timing residuals.
\item We have determined the orbital parameters of
the inferred planet: its orbital period, orbital eccentricity,
semi-major axis projected onto the line of sight, and
mass (with accuracy to within the orbital inclination).
\item Despite the presence of short-period variations
in the timing residuals, due to the nature of these
variations, we were not able to confirm the existence
of a second planet orbiting this pulsar with the period
$P_{2}$ = 3 yrs.
\end{enumerate}

\section{ACKNOWLEDGMENTS}

This work was supported by the Basic Research
Program P-7 of the Presidium of the Russian
Academy of Sciences, subprogram \textquotedblleft Transitional and
Explosive Processes in Astrophysics.\textquotedblright{} This study
would not been possible without the long-term timing
measurements of the pulsar PSR B0329+54 carried
out by T. V. Shabanova.

\appendix

\newpage
\begin{table}
\begin{center}
\setcaptionmargin{0mm} 
\onelinecaptionsfalse
\captionstyle{flushleft} 
\caption{Astrometric and spin parameters of PSR B0329+54}
\medskip
\begin{tabular}{|l|l|}
\hline
 Parameters & Values \\
\hline
RAJ & $03^{h}32^{m}59.373(1)^{s}$  \\
DECJ & $54^{^\circ}34^{\prime}43.49(2)^{\prime\prime}$  \\
PMRA, mas/yr  &  17   \\
PMDEC, mas/yr &  -10 \\
$\nu$, s$^{-1}$ & 1.3995410093399(14) \\
$\dot\nu$, s$^{-2}$ & $-4.011433(2)\times10^{-15}$  \\
$\ddot\nu$, s$^{-3}$ & $3.37(2)\times 10^{-27}$ \\
PEPOCH & 48000        \\
DM, pc/cm$^3$ & 15.4  \\
EPHEM   &  DE405      \\
\hline
\end{tabular}
\footnotetext{RAJ and DECJ are the right ascension and declination, PMRA
	and PMDEC the proper motion in right ascension and declination,
	PEPOCH the epoch to which the coordinates and spin
	frequency of the pulsar are reduced, DM the dispersion measure,
	and EPHEM the ephemerides used to reduce the PATs to the
	barycenter. The numbers in parantheses are the uncertainties in
	the last significant digits.}
\end{center}
\label{tab:par}
\end{table}

\newpage
\begin{table}
\setcaptionmargin{0mm} 
\onelinecaptionsfalse
\captionstyle{flushleft} 
\caption{Parameters of the planet orbiting PSR B0329+54}
\bigskip
\begin{tabular}{|l|l|}
\hline
Parameters & Values \\
\hline
Projected semi-major axis, $a_p\sin i$ (ms)  & 21.58 $\pm$ 0.14
\\
Semi-major axis of relative orbit, $a$ (AU)  & 10.26 $\pm$ 0.07
\\
Period, $P_{1}$ (yr) & 27.76 $\pm$ 0.03   
\\
Eccentricity, $e_{1}$ & 0.236 $\pm$ 0.011   
\\
Mass of the planet, \textit{$m_{c}$}sin\textit{i} ($M_{\oplus}$) & 1.97 $\pm$ 0.19   
\\
\hline
\end{tabular}
\label{tab:planet}
\end{table}  

\newpage
\begin{figure}[h!]
\setcaptionmargin{1mm}
\vbox{\includegraphics[width=1\linewidth]{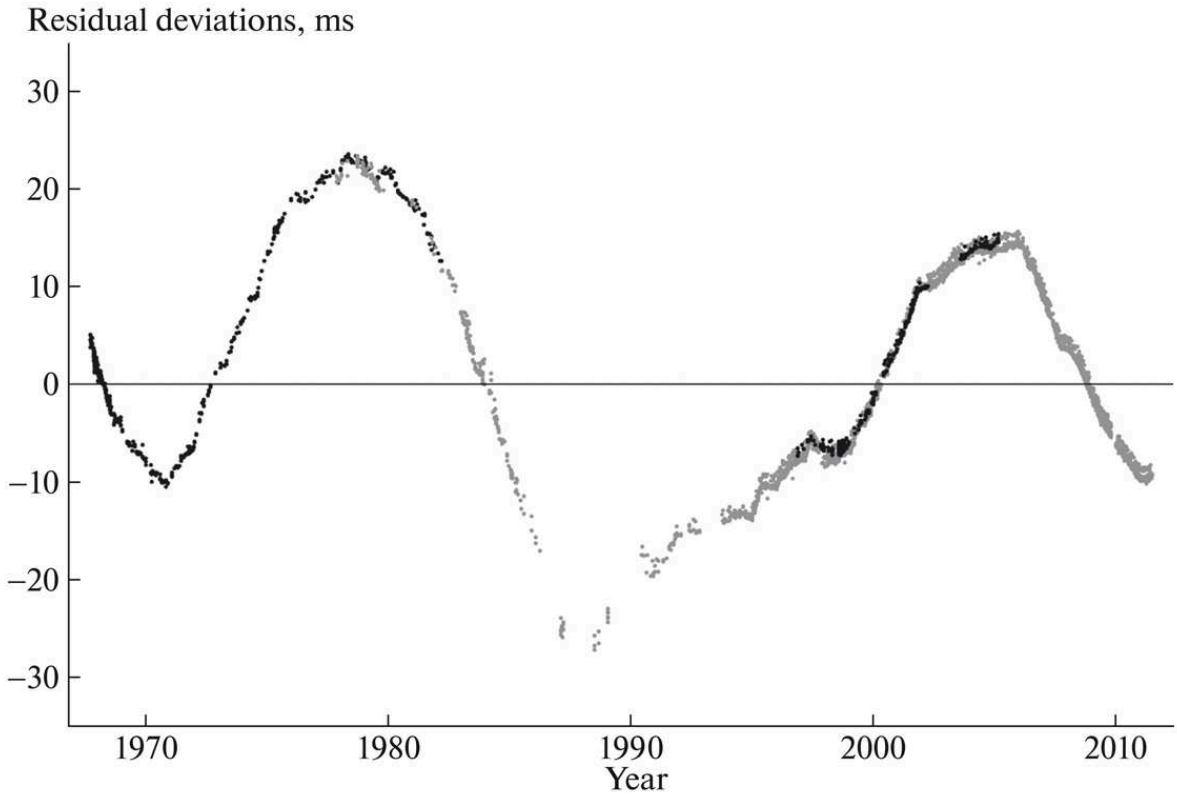}}
\caption{Timing residuals of PSR B0329+54 in 1968--2012. The gray curve shows the residuals corresponding to the PRAO data (102 and 111 MHz) and the black curve the residuals according to the JPL data (2388 MHz, on the left) and KRAO data (610 MHz, on the right).}
\label{ris:fig1}
\end{figure}

\newpage
\begin{figure}[h!]
\setcaptionmargin{1mm}
\vbox{\includegraphics[width=1\linewidth]{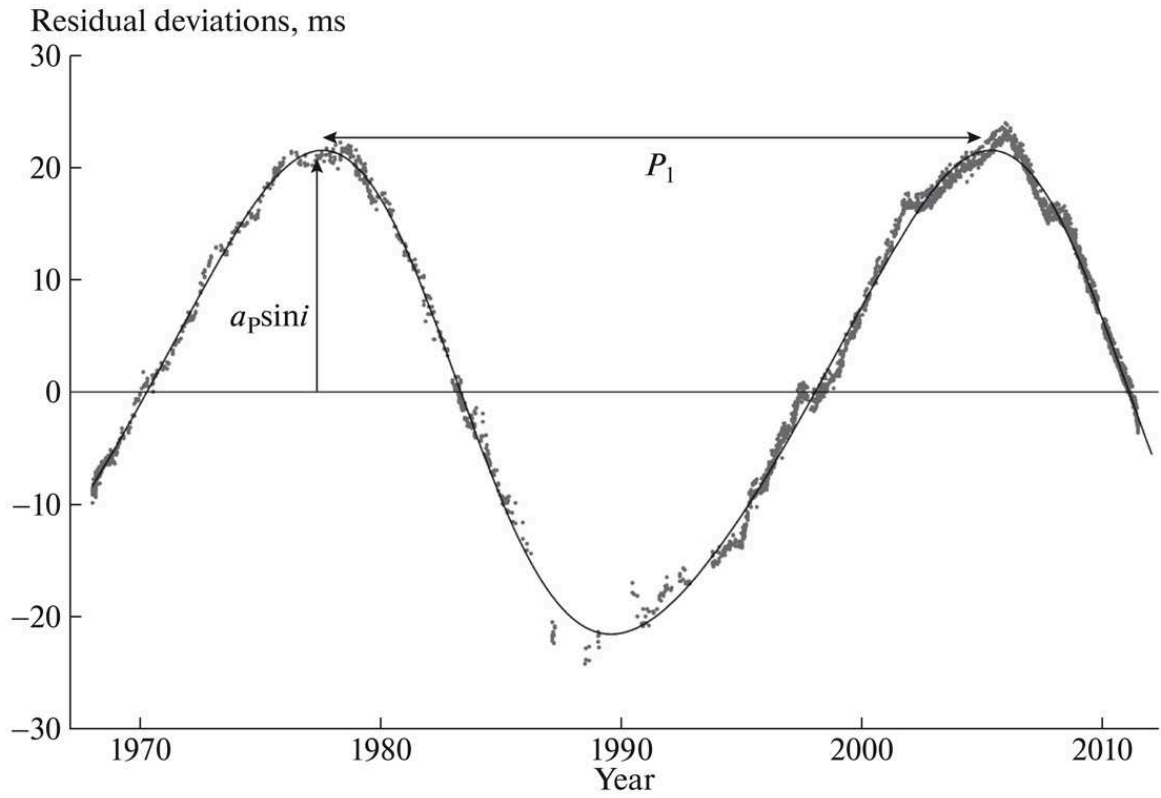}}
\caption{Timing residuals (points) remaining after subtraction of variations of the pulsar spin and a theoretical curve (solid)
	describing the motion of the first planet.}
\label{ris:firg2}
\end{figure}

\newpage
\begin{figure}[h!]
\setcaptionmargin{1mm}
\vbox{\includegraphics[width=1\linewidth]{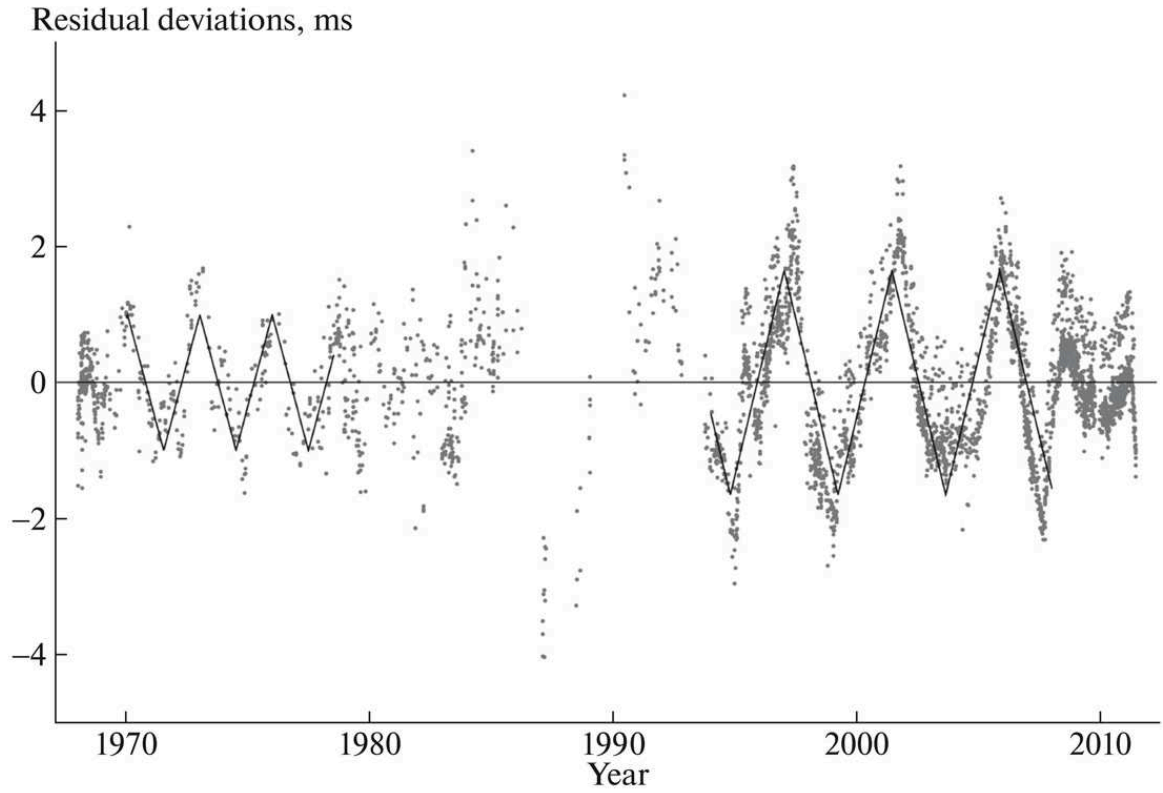}}
\caption{Shape of the timing residuals after subtraction of the model for the motion of the first planet and the intrinsic variations
	of the pulsar spin.}
\label{ris:fig3}
\end{figure}

\end{document}